\documentstyle[amssymb,aps,prl,multicol,epsf]{revtex}

\renewcommand{\narrowtext}{\begin{multicols}{2} \global\columnwidth20.5pc}
\renewcommand{\widetext}{\end{multicols} \global\columnwidth42.5pc}

\newcommand{\ba}{\begin{eqnarray}}
\newcommand{\ea}{\end{eqnarray}}
\newcommand{\be}{\begin{equation}}
\newcommand{\ee}{\end{equation}}

\begin{document}
\draft
\title{Mechanism of pseudogap probed by a local impurity}
\author{Qiang-Hua Wang}
\address{(a) National Laboratory of Solid State Microstructures,%
\\
Institute for Solid State Physics, Nanjing University, Nanjing 210093, China}
\address{(b) Department of Physics, University of California at Berkeley, CA 94720}
\maketitle

\begin{abstract}
The response to a local strong non-magnetic impurity in the
pseudogap phase is examined in two distinctly different scenarios:
phase-fluctuation (PF) of pairing field and d-density-wave (DDW)
order. In the PF scenario, the resonance state is generally
double-peaked near the Fermi level, and is abruptly broadened by
vortex fluctuations slightly above the transition temperature. In
the DDW scenario, the resonance is single-peaked and remains sharp
up to gradual intrinsic thermal broadening, and the resonance
energy is analytically determined to be at minus of the chemical
potential.
\end{abstract}



\narrowtext

Aside from others,\cite{other} two distinctly different scenarios are
proposed for the pseudogap in cuprates,\cite{pgreview} depending on whether
the pseudogap phase is independent of the pairing gap. In the
phase-fluctuation (PF) scenario,\cite{pfscenario} it is speculated that the
normal state contains preformed Cooper pairs, and the phase-fluctuation of
the pairing field destroys superconductivity. As the pairing gap has a
d-wave symmetry in the internal momentum space, the d-wave-like dispersion
of the pseudogap \cite{arpes} follows immediately. An advantage of this
scenario is that it involves no symmetry breaking, and is adiabatically
connected to the paramagnetic Mott insulator. Such a normal state is not a
Fermi liquid. In the second scenario, the normal state is free of pairing
instability, but is in a symmetry-breaking d-density-wave (DDW) state.\cite
{ddwscenario} The latter is an ordered state of staggered orbital current,
and was discussed in other contexts already in the early stage of high-$%
T_{c} $ physics.\cite{affleck} It creates four hole-like Fermi pockets in
the nodal directions. The volume enclosed by the Fermi pockets scales
exactly as the doping level $x$. Thus the pseudogap is from the band
structure effect. The normal state is a Fermi liquid, namely, a DDW metal.

In this Letter we discuss the resonance state due to a strong nonmagnetic
local impurity in the pseudogap phase, which turns out to be markedly
different in these scenarios. In the PF scenario, it is two-peaked near
the Fermi level, broadened by the vortex fluctuations (in addition to the
intrinsic thermal broadening), and is thus strongly temperature dependent
near the superconducting transition temperature $T_c$. (This is a
complementary result to the extended impurity one. \cite{sheehy}) In
contrast, we analytically verify that the resonance state is
single-peaked, remaining sharp and pinned at minus of the chemical
potential in the DDW metal,\cite {jianxin} and identify the underlying
mechanism for the pinning effect. We propose to measure the temperature
and doping dependence of the impurity state by, {\it e.g.}, scanning
tunnelling microscopy,\cite{stm} in order to tell the mechanism of the
pseudogap (or whether the normal state is a Fermi liquid).

In a d-wave superconductor, a local strong impurity is known to give rise to
a resonant state near the Fermi level.\cite{balatsky} The state is almost
real as the scattering rate into the continuum is limited by the vanishing
density of states near the Fermi level (in an unperturbed system) because of
the d-wave pairing symmetry. Qualitatively similar resonant states are found
numerically in a d-wave superconductor with DDW order.\cite{jianxin} This is
because 1) The resonant impurity state {\it near} the Fermi level is a
generic feature of d-wave pairing, and 2) The only significant effect of DDW
is to generate a specific band structure, on top of which d-wave pairing
occurs.

Returning to the normal state, it is natural to expect {\it qualitatively
different} responses to local impurities in the PF and DDW scenarios,
since in the DDW normal state no pairing occurs. An earlier attempt to
address the resonance in the pseudogap phase was made in
Ref.\cite{martin}, but with only limited success for PF. Recently the {\it
extended} impurity was discussed in the PF scenario,\cite{sheehy} and a
numerical study was performed for a {\it local} impurity in the DDW
scenario.\cite{jianxin} In this Letter we compare the behaviors of the
same local impurity in both PF and DDW scenarios.

{\it Phase-fluctuation scenario}: The effective mean field hamiltonian in a
square lattice for a d-wave superconductor may be written as $%
H=\sum_{\langle ij\rangle }(\Psi _{i}^{\dagger }h_{ij}\Psi _{j}+{\rm h.c.}%
)-\mu \sum_{i}\Psi _{i}^{\dagger }\sigma _{3}\Psi_i ,$ where $\Psi
_{i}=(f_{i\uparrow },f_{i\downarrow }^{\dagger })^{{\rm T}}$ is
the Nambu spinor, $\mu $ is the chemical potential, $\sigma _{3}$
is the third Pauli matrix, and
\[
h_{ij}=-t\sigma _{3}+\left(
\begin{array}{cc}
0 & \Delta _{ij} \\
\Delta _{ij}^{\ast } & 0
\end{array}
\right) ,
\]
with $\Delta _{ij}=\Delta _{0}\eta _{ij}\exp (i\varphi _{ij})$, where $%
\Delta _{0}$ is the pairing amplitude, $\eta _{ij}=1$ ($-1$) for $x$%
-direction ($y$-direction) bonds, and $\varphi _{ij}$ is the phase. In the
PF scenario, the pairing field is disordered by thermal and/or quantum
fluctuations of vortices at zero applied magnetic field. In the following
discussion we assume that the vortex fluctuations are thermal.

It is possible to make a singular gauge transform $\Psi _{i}\rightarrow {\rm %
e}^{-i\phi _{i}\sigma _{3}/2}\Psi _{i}$ so that $\Delta _{ij}\rightarrow
\Delta _{0}\eta _{ij}$ no longer carries the phase, whose effect migrates to
the hopping part in $h_{ij}$: $-t\sigma _{3}\rightarrow -t\sigma _{3}{\rm e}%
^{i(\phi _{i}-\phi _{j})\sigma _{3}/2}$. In the continuum limit, the phase
difference between neighboring sites translates to the phase gradient $2{\bf %
q}_{s}\equiv {\rm e}^{-i\phi }\frac{\nabla }{i}{\rm e}^{i\phi }$,
and corresponds to the superfluid velocity. It varies at the
length scale of the London penetration depth, being much larger
than the Fermi wave-length. Thus it is valid to adopt the
quasi-classical approximation, in which Fermions see {\it a
microscopically} constant ${\bf q}_{s}$ while ${\bf q}_{s}$ itself
varies {\it macroscopically}. In this sense, the (matrix) Greens
function $G_{0}$ for the $\Psi $ Fermions influenced by ${\bf
q}_{s}$ is determined by, in momentum space,
\begin{eqnarray}
G_{0}^{-1}({\bf k},i\omega _{n};{\bf q}_{s}) &=&\left(
\begin{array}{cc}
i\omega _{n}-\varepsilon _{{\bf k}+{\bf q}_{s}} & \Delta _{{\bf k}} \\
\Delta _{{\bf k}} & i\omega _{n}+\varepsilon _{{\bf k}-{\bf q}_{s}}
\end{array}
\right)  \nonumber \\
\ \ \ \ \ \ \ \ \ \ \ \ \ \ \ \ \ &\sim &(i\omega _{n}-{\bf q}_{s}\cdot {\bf %
v}_{{\bf k}})\sigma _{0}-\varepsilon _{{\bf k}}\sigma _{3}+\Delta _{{\bf k}%
}\sigma _{1}\label{eq:g01}
\end{eqnarray}
Here $\varepsilon _{{\bf k}}=-2t(\cos k_{x}+\cos k_{y})-\mu $, $\Delta _{%
{\bf k}}=2\Delta _{0}(\cos k_{x}-\cos k_{y})$, ${\bf v}_{{\bf k}}={\bf %
\nabla _{k}}\varepsilon _{{\bf k}}$. As is well known, the low energy
excitations in this system are located around the four nodes ${\bf k}%
_{n=1,2,3,4}=(\pm K,\pm K)$ in the momentum space, with $-4t\cos K=\mu$.
The second line in Eq.(\ref{eq:g01}) is the usual Doppler
approximation.\cite{doppler} At low energies, the Doppler shift ${\bf
q}_s\cdot {\bf v_k}$ can be well approximated by its value at the four
nodes $D_n={\bf q}_s\cdot {\bf v}_{{\bf k}_n}$. We shall use the first
(second) line of Eq.(\ref{eq:g01}) for numerical (qualitative and
analytical) discussion.\cite{caution}

The real space Greens function is obtained by the Fourier transform, $%
G_{0}(i,j,i\omega _{n};{\bf q}_{s})=\sum_{{\bf k}}G_{0}({\bf k},i\omega _{n};%
{\bf q})\exp [i{\bf k}\cdot ({\bf r}_{i}-{\bf r}_{j})]$. Of special interest
is the local Greens function $g(i\omega _{n};{\bf q}_{s})\equiv
G_{0}(i=j,i\omega _{n};{\bf q}_{s})$. Due to the d-wave pairing symmetry, $%
g_{12}=g_{21}=0$. (We suppress the arguments if applicable.) On the other
hand,
\begin{eqnarray}
g_{11}(i\omega _{n};{\bf q}_{s}) &=&-g_{22}^{\ast }=\int dEN_{0}(E;{\bf q}%
_{s})/(i\omega _{n}-E),  \label{Eq:g11} \\
N_{0}(E;{\bf q}_{s}) &\sim &(1/8)\sum_{n=1}^{4}\sum_{\nu =\pm }[|E-\nu
D_{n}|/(8\pi t\Delta _{0})  \nonumber \\
&-&\mu (E-\nu D_{n})^{2}{\rm sgn}(E-\nu D_{n})/\Lambda ^{4}],  \label{Eq:N0q}
\end{eqnarray}
with $\Lambda= 4[\pi t^{3}\Delta _{0}^{3}/(t^{2}+\Delta _{0}^{2})]^{1/4}$.
A cutoff at $|E|>E_c=\min{(4t,4\Delta_0)}$ is necessary in applying
Eq.(\ref{Eq:N0q}). Anticipating its effect in the impurity scattering, we
point out briefly the behavior of $N_0$: 1) It exhibits a four-fold
symmetry in the direction of ${\bf q}_s$. 2) At $|E|\ll E_c$ and $|D_n|\ll
E_c$, the leading contribution comes from the first term in $N_{0}$. It is
particle-hole symmetric around $E=0$ at any $\mu$ and $q_s$. This is the
fundamental property of d-wave pairing between time-reversed particles. 3)
Away from half filling ($\mu\neq 0$), there is a slight asymmetry. To the
first order in $\mu$, this is included in the second term in $N_0$.
Clearly, $N_{0}(E=0;{\bf q}_{s})\propto q_{s}$, as first pointed out by
Volovik. \cite{doppler}

In the presence of a local scattering potential at site $i=0$, the
corresponding Greens function $G$ can be obtained within the T-matrix
approximation (which is exact if the impurity does not spoil the pairing
field),
\begin{eqnarray}
G(i,j,i\omega _{n};{\bf q}_{s}) &=&G_{0}(i,j,i\omega _{n};{\bf q}_{s})
\nonumber \\
+G_{0}(i,0, &i\omega _{n}&;{\bf q}_{s})T(i\omega _{n};{\bf q}%
_{s})G_{0}(0,j,i\omega _{n};{\bf q}_{s}),  \label{eq:g} \\
T^{-1}(i\omega _{n};{\bf q}_{s}) &=&1/(V\sigma _{3}+V_{m}\sigma
_{0})-g(i\omega _{n};{\bf q}_{s}),  \label{eq:tmx}
\end{eqnarray}
where $V$ ($V_{m}$) is the nonmagnetic (magnetic) scattering
strength. To simplify our discussion, we shall consider
nonmagnetic scattering only. With the impurity, the local density
of states (LDOS) is site dependent. At site $i$ it is given by \be
N(i,\omega;{\bf q}_s)=-(1/\pi){\rm Im}G_{11}(i,i,\omega+i0^+;{\bf
q}_s).\label{eq:dosq}\ee  We emphasize that the off-diagonal
elements of $G_{0}$ in Eq.(\ref{eq:g}) contribute to the density
of states. Conceptually, neglecting such contributions, as in
Ref.\cite{martin}, the theory would be hardly related to pairing.

Let us discuss the qualitative behavior of the LDOS in our case. Since $g$
is diagonal, so is the T-matrix. After analytical continuation
$i\omega_n\rightarrow \omega+i0^+$, the resonance in LDOS is determined by
${\rm Re}(T^{-1}_{11})=0$ or ${\rm Re}(T^{-1}_{22})=0$. This is equivalent
to ${\rm Re}[g_{11}(\pm \omega +i0^{+};{\bf q}_{s})] =V^{-1}$, and
immediately implies two resonance peaks in a general situation, an
essential feature of pairing. For a strong scatter, $V^{-1}\rightarrow 0$.
In the case of $q_s=0$, a sharp resonance exists in the LDOS at the sites
nearest to the impurity.\cite{balatsky} The behavior at $q_s\neq 0$ is as
follows: 1) For $\mu =0$ and $V^{-1}=0$, we have perfect particle-hole
symmetry, so that ${\rm Re}[g_{11}(i0^{+};{\bf q}_{s}{\bf )]}=0$ from
Eq.(\ref{Eq:g11}). The resonance is at $\omega=0$. However, {\it it is not
sharp, and its width scales with }$N(0;{\bf q}_{s})$ ($\varpropto q_{s}$)
in accordance with the Fermi golden rule. 2) The effect of a finite $\mu$
and/or $V^{-1}$ is to generate a slight asymmetry, and thus splits the
resonance into two peaks, situated on either side of the Fermi level.
Their widths are identical (different) if $\mu=0$ ($\mu\neq 0$) because of
the behavior of $N_0$. Moreover, at large enough $q_s$ the splitting will
be smeared due to the broadening of both peaks.

Before we proceed, we predict from the above results that even below $T_c$
the resonance may be broadened by an in-plane transport current and/or a
nearby static vortex. Since $N_0$ is four-fold symmetric in the direction
of the relevant super-current, so is the broadening phenomenon.

Thermal phase fluctuations are governed by the Kosterlitz-Thouless (KT)
theory. \cite{kttheory} In the quasi-classical approximation, this just
amounts to an average over ${\bf q}_{s}$. For illustrative purpose, one
can assume a Gaussian distribution $\exp (-{\bf q}_{s}^{2}/2n_{v})/(2\pi
n_v)$ for ${\bf q}_s$, with $n_{v}$ scaling with the density of free
vortices (anti-vortices).\cite{sheehy} The average LDOS is thus $
N(i,\omega) =\langle N(i,\omega;{\bf q}_s)\rangle$. The averaging makes it
inconvenient to determine the resonance exactly, since it is the Greens
function that should be averaged instead of the T-matrix alone. However,
qualitative features of the resonance in the LDOS are roughly the same as
discussed above, but with a characteristic scale of $q_{s}$ given by
$\sqrt{n_{v}}$, the inverse vortex spacing.

\widetext
\begin{figure}[tbp]
\hskip 1cm \epsfxsize=15cm\epsfbox{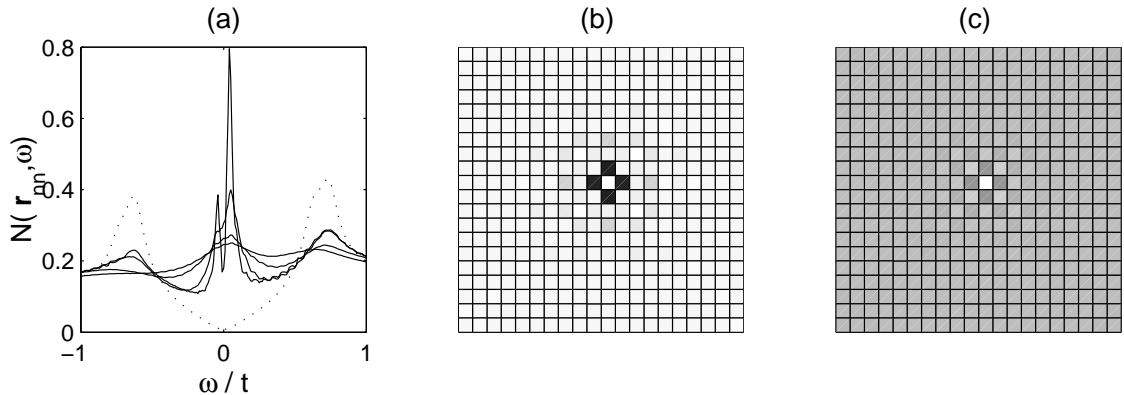}
\caption{ Results with $\Delta _{0}=0.17t$, $\protect\mu =-0.3t$ and $V=100t$%
. (a) $N({\bf r}_{nn},\protect\omega )$ versus $\protect\omega $.
Solid lines: $n_v=0\sim 10^{-6},10^{-4},10^{-3}$, and $5\times
10^{-3}$ with decreasing peaks. The dotted line is the LDOS at
$n_v=0$ and $V=0$ for comparison. (b) $N({\bf r},0.05t)$ at
$n_v=0$. The impurity is at the center. (c) The same as (b) for
$n_v=5\times 10^{-3}$. The gray scale is the same in (b) and (c).}
\end{figure}

\narrowtext

All of the above predictions are indeed seen in our numerical results. We
use Eqs.(\ref{eq:g01}) (the first line) and (\ref{eq:g})-(\ref{eq:dosq}),
and average over ${\bf q}_s$. In Figs.1 we present results for $\Delta
_{0}=0.17t$, $\mu =-0.3t$ and $V=100t$. (The results do not change much in
the unitary limit $V\gg t$.) Fig.1(a) plots the LDOS at ${\bf r}_{nn}$
nearest to the impurity as a function of energy. For $%
n_{v}\leq 10^{-6}$ (or vortex spacing $d_{v}\geq 10^{3}$ in units
of crystal lattice constant $a_0$), the resonance is sharp and
indistinguishable from that with no vortices at all.
However, it begins to degrade at $n_{v}\geq 10^{-4}$ ($d_{v}\leq 10^{2}a_{0}$%
), and becomes almost featureless at $n_{v}=5\times 10^{-3}$ ($d_{v}=14a_{0})
$. Translating to the temperature dependence using the KT expression $%
n_{v}\sim \exp [-\sqrt{aT_{c}/(T-T_{c})}]$ (with $a=5$ for estimation),
one expects no significant change of the resonance at $T-T_c<0.07 T_{c}$,
but it is suddenly degraded as soon as $T-T_c>0.1T_c$. For $T_{c}=40K$ as
in a typical underdoped cuprate, the temperature window for this
phenomenon to happen is within $4K$. While the exact number should not be
taken seriously, {\it the sudden degrading of the resonant impurity state
is a robust and peculiar feature of the PF scenario}. A similar case was
found for an extended impurity elsewhere.\cite{sheehy} The dotted line in
Fig.1(a) is the DOS at $n_{v}=0$ and $V=0$ for comparison. It also shows
the slight doping-induced particle-hole asymmetry. Fig.1(b) shows the
spatial distribution of LDOS at $\omega=0.05t$ ($\ll |\mu|$), one of the
resonance energies, when $n_{v}=0$,\cite{balatsky} which should be
compared to Fig.1(c) at $n_{v}=5\times 10^{-3}$, upon which the contrast
for the four-fold structure is much weaker.

{\it DDW scenario}: We assume that the effective Hamiltonian for the uniform
DDW metal is\cite{ddwscenario}

\begin{equation}
H=-\sum_{\langle ij\rangle }[(t+iD_{ij})c_{i}^{\dagger }c_{j}+{\rm h.c.}%
]-\sum_{i}\mu c_{i}^{\dagger }c_{i},
\end{equation}
where spin index is suppressed, $D_{ij}=D_0\eta_{ij}(-1)^{i}$ with $D_0$
being the DDW order parameter.

It proves useful to introduce two sublattices A and B, and denote $c_{i\in
A}=A_{i}$ and $c_{i\in B}=B_{i}$. The real-space sublattice Greens
function is

\begin{equation}
\left(
\begin{array}{cc}
G_{AA}^{(0)} & G_{AB}^{(0)} \\
G_{BA}^{(0)} & G_{BB}^{(0)}
\end{array}
\right) =\frac{1}{2}\sum_{{\bf k},\nu =\pm }\frac{A_{{\bf k},\nu } \exp (i%
{\bf k}\cdot {\bf r}_{ij})}{i\omega _{n}+\mu -\nu E_{{\bf k}}},
\end{equation}
where $A_{{\bf k},\nu }=\sigma _{0}+\nu \sigma _{1}X_{{\bf k}}/E_{{\bf k}%
}+\nu \sigma _{2}D_{{\bf k}}/E_{{\bf k}},$ $E_{{\bf k}}=\sqrt{X_{{\bf k}%
}^2+D_{{\bf k}}^2}$, $X_{{\bf k}}=2t(\cos k_{x}+\cos k_{y})$, and $D_{{\bf k}%
}=2D_0(\cos k_{x}-\cos k_{y})$. Note that the summation over the momentum
space is limited to the reduced Brillouin zone. In order to study the
impurity problem, we need the real-space Greens
function $G_{c}^{(0)}(i,j,i\omega _{n})$ in terms of the original $c$%
-electrons. This is related to the above as \be G_{c}^{(0)}(i\in \alpha
,j\in \beta ,i\omega _{n})=G_{\alpha \beta }^{(0)}(i,j,i\omega _{n}),\ee
where $\alpha ,\beta =A,B.$

The unperturbed on-site Greens function is independent of sublattices,
$g_c(i\omega_n)\equiv G^{(0)}_c(0,0,i\omega_n)=\int dE
N_0(E)/(i\omega_n-E)$ with $N_{0}(E)=\sum_{{\bf k}\nu =\pm }\delta (E-\nu
E_{k}+\mu )\sim |E+\mu |/(8\pi Dt)$ being the unperturbed DOS. The second
equality in $N_0$ requires a cutoff at $|E+\mu|>E_{c}=\min (4D,4t)$. The
symmetry around $E=-\mu$ in $N_0$, instead of at $E=0$ in the case of
pairing, is exact in our model. This is because doping the system does not
change the two {\it symmetric} bands generated by DDW, but just shift the
Fermi level.

Now, the Greens function in the presence of the impurity can again
be obtained within the T-matrix formulation,
\begin{eqnarray}
G_{c}(i,j,i\omega _{n}) &=&G_{c}^{(0)}(i,j,i\omega _{n})  \nonumber \\
&&+G_{c}^{(0)}(i,0,i\omega _{n})T(i\omega _{n})G_{c}^{(0)}(0,j,i\omega _{n}),
\label{eq:gc} \\
T^{-1}(i\omega _{n}) &=&V^{-1}-g_c(i\omega_n).
\end{eqnarray}
The resonance state is defined by ${\rm Re}[T^{-1}(\omega+i0^+)]=0$, or
equivalently ${\rm Re}[g_c(\omega+i0^+)]=V^{-1}$. Using the approximate
$N_0$, $g_c(\omega+i0^+)$ is given by \be -{\rm
sgn}(\omega+\mu)N_0(\omega)\ln [\frac{E_{c}^{2}}{(\omega +\mu
)^{2}}-1]-i\pi N_0(\omega).\ee Thus the resonance occurs below (above)
$-\mu $ for a {\it finite} positive (negative) potential $V$. For a
nonmagnetic impurity, it is single-peaked because of the unique condition
for the resonance to occur. Again the energy width of the resonance scales
with $N_0$. In the unitary limit $V^{-1}\rightarrow 0$, the resonance
energy is $\omega =-\mu$ from the above expression of $g_c$ and
$N_0(-\mu)=0$. In fact this result is {\it exact} since the exact symmetry
in $N_0$ mentioned above guarantees ${\rm Re}[g_c(-\mu+i0^+)]=0$. This
resonance energy is exactly at the mid-point of the two symmetric bands,
in much the same way as the mid-gap state exists in a semiconductor.
Furthermore it would be infinitely sharp since $N_{0}(-\mu)=0$.

The DDW order should not fluctuate significantly once it is well
developed, because it is an Ising-like order parameter so that no
Goldstone mode exists. Therefore the thermal rounding of the
resonance is gradual, with no abrupt change just above $T_c$, in
contrast to the case in the PF scenario.

\begin{figure}[tbp]
\epsfxsize=8.5cm\epsfbox{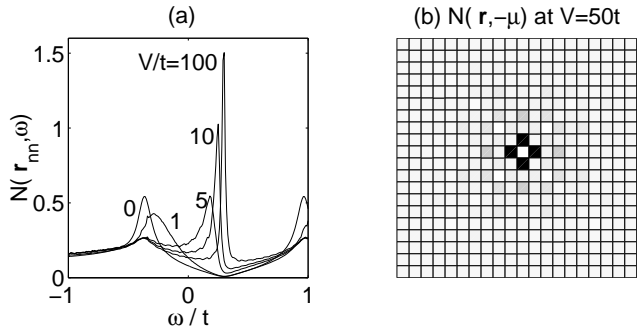}\vskip 0.15cm
\caption{Results with $D_0=0.17t$ and $\protect\mu=-0.3t$. (a) $N({\bf r_{nn}%
},\protect\omega)$ versus $\protect\omega$. (b) $N({\bf r},\protect\omega=-%
\protect\mu)$ versus ${\bf r}$.}
\end{figure}

The LDOS $N(i,\omega )$ can be easily calculated from
Eq.(\ref{eq:gc}). In Fig.2(a) we present the LDOS at ${\bf
r}_{nn}$ nearest to the impurity. The resonance is single-peaked
and robust as long as $V\gg t,D$. In Fig.2(b) we present the
spatial dependence of the LDOS at the resonance energy. The
pattern is also four-fold symmetric, similar to the case of a
d-wave superconductor except that the resonance here is at $-\mu $
instead of at zero energy.

After the submission of this paper, we become aware of a related but
independent work.\cite{morr}

{\bf Acknowledgment} Useful discussions with Dung-Hai Lee, A. V. Balatsky,
J. C. Davis and Jian-Xin Zhu are appreciated. This work is supported by the
National Natural Science Foundation of China and the Ministry of Science and
Technology of China ({NKBRSF-G1999064602}), and in part by the Berkeley
Scholars Program.

\widetext

\end{document}